\documentclass[pra,nofootinbib]{revtex4}

\usepackage{amsmath}
\usepackage{graphicx}
\usepackage{epsfig}
\def\>{\rangle}
\def\<{\langle}
\def\be{\begin{equation}}
\def\ee{\end{equation}}
\def\bea{\begin{eqnarray}}
\def\eea{\end{eqnarray}}
\def\mn1#1{-#1}

\begin{document}


\title{Constructing physically intuitive graph invariants}
\author{Terry Rudolph}
\affiliation{Bell Labs, 600-700 Mountain Ave., Murray Hill, NJ 07974,
U.S.A.}
\affiliation{rudolpht@bell-labs.com}

\date{June 9th, 2002 (my birthday!)}       
\begin{abstract}
In this brief note I try to give a simple example of where
physical intuition about a collection of interacting qubits can
lead to the construction of ``natural'' versions of what are,
generically, quite abstract mathematical objects - in this case
graph invariants.
\end{abstract}

\maketitle

By a \emph{graph}, a mathematician generally means a collection of
points (\emph{vertices}) and a list indicating how they are
connected (a collection of \emph{edges}). The study of graphs and
their properties is a huge industry, with applications from the
completely abstract (e.g. classification of algebras) to the very
applied (e.g. network routing). The simplest type of graph, and
the only type which I'll consider here, is one that has either
zero or one (no multiple) undirected connections between any two
vertices.

Given two graphs, such as in Fig.~1., one of the simplest
questions to ask is whether they are actually the same graph; this
is known as the \emph{graph isomorphism} problem.  If the two
graphs are different, the question is often simple to decide. For
example, the two graphs may have different numbers of vertices,
although these two do not. If they have the same number of
vertices they may have a different number of edges, although
again these two do not. If they have the same number of vertices
and edges, it may be, as in Fig.1., that only one of the graphs
has a vertex which is connected to exactly four other vertices,
indicating clearly the graphs cannot be isomorphic. More
generally, we can list the
\emph{degrees} of each graph's vertices and check if the lists
are identical. (Note that we are only interested in the underlying
connectivity of the graph, and so the distances between vertices
are not important; the same graph can be drawn many different
ways.)

Once we have performed these few simple checks, which I should
emphasize are capable only of telling us whether the graphs are
different, things get a little trickier. To be convinced that two
graphs are the same we need to find a map of the vertices of
graph~2 to those of graph~1. That is, we try and find a
relabelling of the second graph's vertices such that it is now
manifestly clear that it is the same as the first graph. The
problem is that the number of ways we can re-label the
$N$ vertices of a graph is
$N!$ -- the number of permutations of $N$ items. This amounts to a
very large number of possible relabellings, and for modest values
of $N$ searching through them all becomes computationally
infeasible. (Of course if someone magically hands us the correct
relabelling it is very easy to check that the two graphs are the
same!)
\begin{figure}[h]
\begin{center}
\includegraphics[width=80mm]{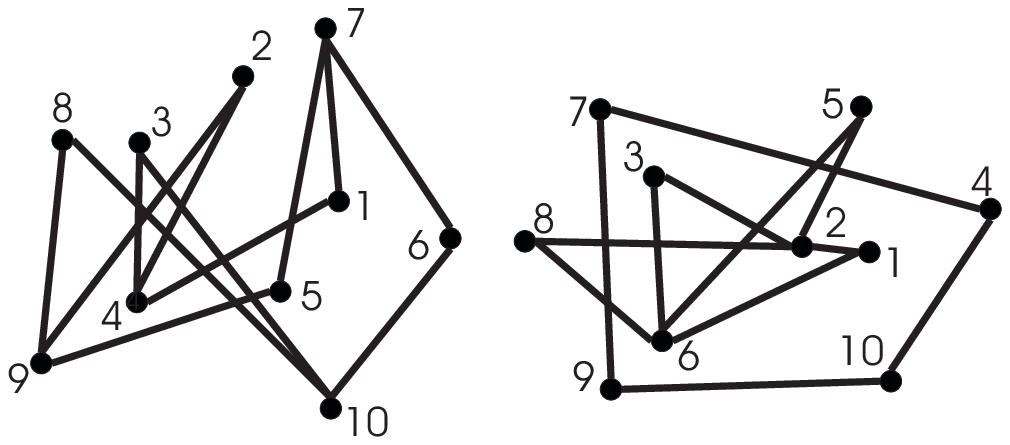}
\end{center}
\centerline{Fig.1. Two non-isomorphic ten vertex graphs}
\end{figure}

In this note I'll be considering \emph{graph invariants}. These
are properties of the graph, such as number of vertices, number of
edges or degrees of vertices mentioned above, which are relatively
easy to compute and which
\emph{must} be the same if the two graphs are the same.
That is, if we compute the graph invariants for two graphs and
they are different we know for sure that the graphs are
different; if they're the same we have learned nothing.

A common way of encoding a graph is through its adjacency matrix
- a real symmetric matrix with a 1 at position $(i,j)$ if vertex
$i$ is connected to vertex $j$ and a zero otherwise. For example
the adjacency matrices of the two graphs in Fig.2. are
\begin{equation}
 A=\left[
{\begin{array}{rrrrr}
0 & 0 & 0 & 0 & 1 \\
0 & 0 & 0 & 0 & 1 \\
0 & 0 & 0 & 0 & 1 \\
0 & 0 & 0 & 0 & 1 \\
1 & 1 & 1 & 1 & 0
\end{array}}
 \right],\;
 B= \left[
{\begin{array}{rrrrr}
0 & 1 & 0 & 1 & 0 \\
1 & 0 & 1 & 0 & 0 \\
0 & 1 & 0 & 1 & 0 \\
1 & 0 & 1 & 0 & 0 \\
0 & 0 & 0 & 0 & 0
\end{array}}
 \right]
\end{equation}
\begin{figure}[h]
\begin{center}
\includegraphics[width=60mm]{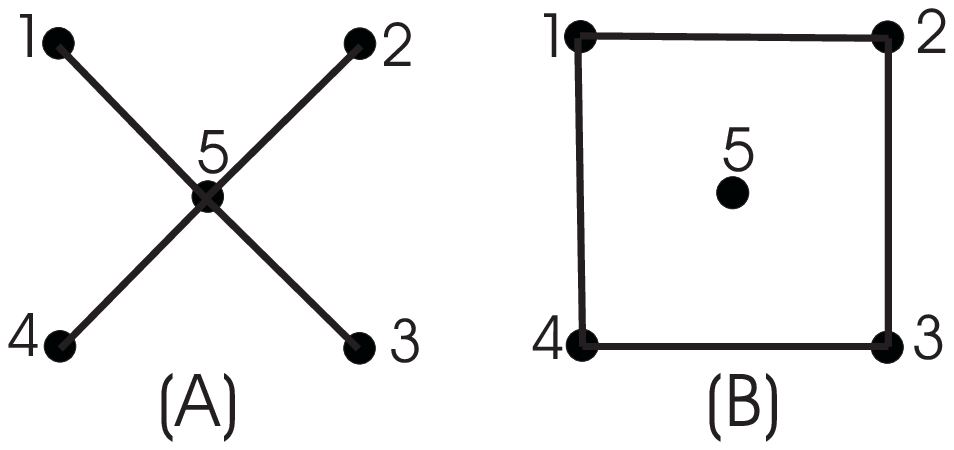}
\end{center}
\centerline{Fig.2. Two graphs with the same adjacency matrix eigenvalues.}
\end{figure}

Given the adjacency matrices $G_1$, $G_2$ of two graphs, we
rephrase the problem of deciding whether the two graphs are
isomorphic, as the question as to whether there exists a
permutation matrix
$\sigma$ such that $\sigma^T G_2 \sigma=G_1$. A more
sophisticated graph invariant than those mentioned above consists
of the eigenvalues of the adjacency matrix. That is, if the
eigenvalues of $G_2$ are different from those of $G_1$, then $G_1$
and $G_2$ are definitely \emph{not} isomorphic. This is because
the eigenvalues of $\sigma^T G_2 \sigma$ are solutions to the
equation $\det(\sigma^T G_2 \sigma-\lambda I)=0$. Since
$\sigma^T \sigma=I,$ and $\det(AB)=\det(BA)$, this becomes $\det(\sigma^T (G_2 -\lambda
I)\sigma)=\det(\sigma\sigma^T (G_2 -\lambda I))=\det(G_2 -\lambda
I)$; the latter is the eigenvalue equation for $G_2$. Hence
applying a permutation to a matrix doesn't change its eigenvalue
equation.

\emph{Spectral graph theory} is the area of mathematics devoted to
analyzing graphs through the eigenvalue spectra of their adjacency
matrix; whole books have been written on the subject [1]. As with
any graph invariants however, there exist annoying non-isomorphic
graphs which are not distinguished by them. The two 5-vertex
graphs $A$ and $B$ of Fig.~2 and Eq.~(1). are an example, they
both have the eigenvalues
$\{-2,[0]^3,2\}$. (Obviously these two graphs could be
distinguished by other means!) What I will explain later on, is
that if we have a 5-dimensional adjacency matrix
$G$, and from its elements we construct the 10-dimensional symmetric matrix
\begin{equation}G^{(2)}\equiv
\left(\begin{matrix}
  0 & G_{23} & G_{24} & G_{25} & G_{13} & G_{14} & G_{15} & 0 & 0 & 0 \\
  G_{23} & 0 & G_{34} & G_{35} & G_{12} & 0 & 0 & G_{14} & G_{15} & 0 \\
  G_{24} & G_{34} & 0 & G_{45} & 0 & G_{12} & 0 & G_{13} & 0 & G_{15} \\
  G_{25} & G_{35} & G_{45} & 0 & 0 & 0 & G_{12} & 0 & G_{13} & G_{14} \\
  G_{13} & G_{12} & 0 & 0 & 0 & G_{34} & G_{35} & G_{24} & G_{25} & 0 \\
  G_{14} & 0 & G_{12} & 0 & G_{34} & 0 & G_{45} & G_{23} & 0 & G_{25} \\
  G_{15} & 0 & 0 & G_{12} & G_{35} & G_{45} & 0 & 0 & G_{23} & G_{24} \\
  0 & G_{14} & G_{13} & 0 & G_{24} & G_{23} & 0 & 0 & G_{45} & G_{35} \\
  0 & G_{15} & 0 & G_{13} & G_{25} & 0 & G_{23} & G_{45} & 0 & G_{34} \\
  0 & 0 & G_{15} & G_{14} & 0 & G_{25} & G_{24} & G_{35} & G_{34} & 0
\end{matrix}\right),
\end{equation}
\emph{then the eigenvalues of this larger matrix are a graph
invariant}, and in fact are a more powerful invariant than those
of the original matrix $G$. If we take the two adjacency matrices
$A,B$ of Eq.~(1) and use Eq.~(2) to construct $A^{(2)}$ and
$B^{(2)}$, what I will call their
\emph{level 2} matrices, then we find that the eigenvalues of
$A^{(2)}$ are $\{-\sqrt{6},[-\sqrt{2}]^3,[0]^2,[\sqrt{2}]^3,\sqrt{6}\}$ while
those of $B^{(2)}$ are $\{-2\sqrt{2},-2,[0]^6,2,2\sqrt{2}\}$.
 The fact these eigenvalues are different proves the non-isomorphism
of the two graphs\footnote{ It is interesting to note that, since
the
$G_{ij}$'s are all equal to 1 or 0, the level 2 matrix of Eq.~(2)
is itself the adjacency matrix of a graph. The graphs of Fig.~1.
are in fact the graphs of $A^{(2)}$ and  $B^{(2)}$. Note that
this means we can use  the whole machinery already developed for
the spectral graph theory of adjacency matrices for analysing
these higher level matrices.}

 All this is not
particularly useful unless we know how to find level 2 matrices
for graphs with more than 5 vertices! It turns out we can, and
they are always ${N \choose 2} \times {N \choose 2}$ dimensional.
To construct a level 2 matrix for a graph of an arbitrary number
of vertices we follow this procedure. We first define an
$(N-1)\times N$ dimensional indexing matrix ${\cal I}_N$ which
contains the entries
$1,2,\ldots,{N\choose 2}$ arranged as this example for $N=6$
indicates:
\begin{equation}
{\cal I}_6=\left(
\begin{matrix}
  0 & 1 & 2 & 3 & 4 & 5 \\
  0 & 0 & 6 & 7 & 8 & 9 \\
  0 & 0 & 0 & 10 & 11 & 12 \\
  0 & 0 & 0 & 0 & 13 & 14 \\
  0 & 0 & 0 & 0 & 0 & 15
\end{matrix}\right)
\end{equation}

Using this indexing matrix we define two functions: $\alpha(i)$
is the \emph{row} of ${\cal I}_N$ which contains the integer
$i$, while $\beta(i)$ is the \emph{column} of ${\cal I}_N$ which
contains $i$. The $(i,j)$'th element of the level 2 matrix for a
graph $G$ is then given by

\begin{eqnarray}
G^{(2)}_{ij}&=&\delta_{\alpha(i)\alpha(j)}G_{\beta(i)\beta(j)}
+\delta_{\alpha(i)\beta(j)}G_{\beta(i)\alpha(j)}\nonumber\\
&&+\delta_{\beta(i)\alpha(j)}G_{\alpha(i)\beta(j)}
+\delta_{\beta(i)\beta(j)}G_{\alpha(i)\alpha(j)},
\end{eqnarray}
where $\delta$ denotes the usual kronecker delta function.

This all seems, on the face of it, a little unlikely. So now for
the physics which underlies the construction of the $G^{(2)}$
matrices, and which in fact shows us how to construct level $n$
matrices $G^{(n)}$ (with each value of $n$ giving a stronger
invariant than the preceding one) up to level
$n=\lfloor\frac{N}{2}\rfloor$.

We begin by considering $N$ interacting qubits (two-level atoms
say), each of ground state $|0\>$ and excited state
$|1\>$ with transition frequency $\omega_0$.
Assume they are interacting via an (excitation-)exchange
Hamiltonian, but that qubit
$i$ only interacts with qubit $j$ if vertices $i$ and $j$ are
connected in the graph $G$. The generic interaction Hamiltonian is
of the form:
\begin{equation}\label{int}
H_{\text{int}}(G)= g \sum_{ i\sim j}
\left(S^+_iS^-_j + S^-_iS^+_j\right),
\end{equation}
Here $S_i^+=|1\>\<0|$ ($S_i^-=|0\>\<1|$) is the raising(lowering)
operator for qubit $i$, and $i\sim j$ means vertex $i$ is
connected to vertex $j$ in
$G$. $g$ is a coupling constant, which we take to be equal for all
interacting qubits (the ``small sample'' or ``long wavelength''
limit for atomic spectroscopists), and we normalize it to 1.

The number of possible states of the $N$ qubits is $2^N$. For our
purposes we will label the states by which qubits are excited (in
the state $|1\>$). For example the 5 qubit state
$|0\>\otimes|1\>\otimes|0\>\otimes|0\>\otimes|1\>$ will be labelled simply as
$|25\>$, which is a ``bi-exciton'' state. The nature of the Hamiltonian
(\ref{int}) is such that its matrix elements between states with
a different number of excited qubits are always 0; e.g.
$\<23|H_{\text{int}}(G)|134\>=0$ regardless of $N$. This is
because the interaction conserves excitation - if one qubit goes
``up'', the other must come ``down''. Furthermore, it is clear
from the form of (\ref{int}), that even if the two states have the
same number of excited qubits, the matrix element of
$H_{\text{int}}(G)$ can only be non-zero if the two states have
a different amount of excitation in one, and only one, \emph{pair}
of qubits.\footnote{In general $H_{\text{int}}(G)$ is block
diagonal, with each block corresponding to a different exciton
level (number of excited qubits) $n$. Each block is a ${N \choose
n} \times {N \choose n}$ matrix whose eigenvalues form a graph
invariant. In abstract terms, the level $n$ matrix $G^{(n)}$ can
be constructed by indexing each row and each column of the
matrix  by an
$n$-tuple of integers chosen from $1,\ldots,N$. Denote the $n$-tuple
corresponding to row/column $i$ by $S_i$.  The element
$G^{(n)}_{i,j}$ can be nonzero only if the set $S_i\cup S_j\setminus S_i\cap S_j$ contains
exactly $2$ integers, say $a,b$. If it does, then
$G^{(n)}_{i,j}=G_{a,b}$.}

Consider the subset of states which contain only one excited
qubit:
$|1\>,|2\>,\ldots,|N\>$. If we write the matrix for
$H_{\text{int}}(G)$ using this subset of states, we find
precisely the adjacency matrix of $G$. The eigenvalues of this
matrix tell us at which energies we would see absorption lines if
we shone appropriate light at our qubit cluster. (Since we have
considered just the interaction Hamiltonian, these eigenvalues
are strictly speaking the \emph{shifts} from
$\omega_0$ of the energy levels). Now, thinking physically, its
obvious that the order in which some human experimenter chooses to
label the atoms is irrelevant to the energy shifts she will see.
In other words, these energy shifts form a graph invariant (as we
know).

The level 2 matrices discussed above are simply the matrices for
what a spectroscopist would call the ``bi-exciton'' energy
manifold. That is, we look at the subset of states for which two
atoms are excited:
$|12\>,|13\>,\ldots,|1N\>,|23\>,\ldots,|2N\>,\ldots,|(N-1)N\>$.
These form the second submatrix of $H_{\text{int}}(G)$. Once again
it is physically obvious that the way we relabel the atoms cannot
affect the physical properties we observe, and thus the
eigenvalues of this matrix also form a graph invariant.

The main purpose of this short note is to point out that
\emph{any} physical quantity we can compute for this hypothetical
cluster of qubits, for which its intuitively obvious the labelling
order is irrelevant, forms a graph invariant. This includes the
emission spectrum, the absorption spectrum, total transition
rates from a given exciton manifold to the one below it, the
average amount of pairwise entanglement when the atoms sit in a
thermal bath, and so on. What is interesting about these other
possibilities, is that they naturally incorporate the
\emph{eigenvectors} of the adjacency matrix, something which from
my (admittedly very brief) reading of the graph theory literature
appear to be somewhat underexploited.

%

Finally I should mention that graph theorists sometimes use a
slightly different matrix for encoding graphs, known as the
\emph{Laplacian} matrix. This is essentially the adjacency matrix
on the off-diagonal elements, but now with non-zero diagonal
elements whose magnitudes reflect the degree of that vertex.
Studying the eigenvalues and other properties of these matrices
seems to also be quite popular. In the quantum mechanical picture
I'm advocating here, the Laplacian matrix arises when we use not
just the interaction Hamiltonian, but an appropriately chosen
``free'' Hamiltonian as well. Similar constructions to those
above yield ``higher level'' Laplacian matrices, and once again
they can form stronger invariants than the simple, level one,
standard Laplacian matrix.

I expect that in fact physicists stand to gain more from learning
some spectral graph theory, than mathematicians stand gain from
learning some physics. For example, the automorphism groups of a
graph, which are much studied by mathematicians, would seem to be
directly related to certain optical properties of the emission
spectra. However the purpose of this note is to give physicists
some feeling of how physical intuition can lead to natural
invariants of abstract mathematical structures. The most famous
(but vastly more complicated) example of this is of course Ed
Witten's construction of knot invariants by considering an
appropriate conformal field theory.
\\*
\begin{center}\textbf{References}\end{center}

[1] The two resources I have used are: Chapter~1 of Edwin van
Dam's thesis at http://cwis.kub.nl/
${}^\sim$few5/center/staff/dam/pub.htm; and Chapter~1 of the
book {\it Spectral Graph Theory} by Fan Chung, which is available
at: http://math.ucsd.edu/${}^\sim$fan/cbms.pdf


\begin{center}\textbf{An afterword for graph
theorists}\end{center}
To test the ideas discussed above I chose two ``very similar'' 24
vertex, strongly regular graphs with identical adjacency (and
Laplace) matrix spectra. They are contained in a list available at
{http://www.maths.gla.ac.uk/${}^\sim$es/reggraph.html} in the file
``I.graphs.24.gz''. The two graphs I chose have four distinct
eigenvalues of their adjacency matrices. The graphs are:

\begin{center}{ Spectrum $=\{[8]^1,[2]^{11},[-2]^{9},[-4]^{3}\}$}
\\
1.$FF0001E3800003F00030F001998019FCC03A30C21E1E01CC303C1542A892610A52592$
\\
$|Aut|$ =   384 (1,2,7,8,9,10,11,12)(3,4,5,6,13,14,15,16,17,18,19,20,21,22,23,24)
\\
2.$FF0001E3800003F00030F001998019FCC03A30C21E1E01CC303C1542A862911231692$
\\
$|Aut|$ =   384  (1,2,7,8,9,10,11,12)(3,4,5,6,13,14,15,16,17,18,19,20,21,22,23,24)
\end{center}
(The adjacency matrix is found by converting the hexadecimal
number to binary, this binary string is a concatenation of the
upper triangular part of the adjacency matrix. I have no idea
about the notation used for the automorphism group.)
\\*
 It turns out that the level \emph{two} matrices of these graphs still have
identical spectra, however at level \emph{three} the spectra are
different, showing the graphs to not be isomorphic. Note that in
general the level
$k$ matrix is $O(N^k\times N^k)$, which according to complexity theorists
is not too bad. However, I assure you it means programming skills
far exceeding mine are required to play around with any much
larger examples.

\end{document}